\documentclass[aps,prb,twocolumn,showpacs,amsmath,amssymb]{revtex4-1}
\usepackage{graphicx,epsf}
\usepackage{bm}

\newcommand{\be}{\begin{equation}}
\newcommand{\ee}{\end{equation}}
\newcommand{\bea}{\begin{eqnarray}}
\newcommand{\eea}{\end{eqnarray}}
\newcommand{\up}{\uparrow}
\newcommand{\down}{\downarrow}
\newcommand{\bwt}{\begin{widetext}}
\newcommand{\ewt}{\end{widetext}}
\newcommand{\ham}{\mathcal{H}}

\newcommand{\ra}{\rangle}
\newcommand{\la}{\langle}
\newcommand{\bsb}{\begin{subarray}}
\newcommand{\esb}{\end{subarray}}
\newcommand{\largem}{\!\!}

\newcommand{\eins}{\mbox{$1 \hspace{-1.0mm} {\bf l}$}}

\newcommand{\vecv}[4]{
\left(\largem
 \begin{tabular}{c}
  $#1$ \\
  $#2$ \\
  $#3$ \\
  $#4$
  \end{tabular}
  \largem
\right)
}

\newcommand{\vech}[4]{
\left(\largem
 \begin{tabular}{c}
  $#1$ \! $#2$ \! $#3$ \! $#4$
  \end{tabular}
  \largem
\right)
}

\newcommand{\mat}[4]{
\left(
\largem
 \begin{tabular}{cc}
  $#1$ & $#2$ \\
  $#3$ & $#4$
  \end{tabular}
  \largem
\right)
}

\begin{document}

\title{Tight-binding study of bilayer graphene Josephson junctions}

\author{W.A. Mu\~noz}
\author{L. Covaci}
\author{F.M. Peeters}
\affiliation{Department Fysica, Universiteit Antwerpen, Groenenborgerlaan 171, B-2020 Antwerpen, Belgium}

\date{\today}

\begin{abstract}
Using highly efficient simulations of the tight-binding Bogoliubov-de Gennes model we solved self-consistently for the  pair
correlation and the Josephson current in a Superconducting-Bilayer graphene-Superconducting Josephson junction.
Different doping levels for the non-superconducting link are considered in the short and long junction regime.
Self-consistent results for the pair correlation and superconducting current resemble those reported previously for single layer
graphene except in the Dirac point where remarkable differences in the proximity effect are found as well as a suppression of the
superconducting current in long junction regime. Inversion symmetry is broken by considering a potential difference
between the layers and we found that the supercurrent can be switched if  junction length is larger than the Fermi length.
\end{abstract}



\maketitle


\section{Introduction}

Ever since graphene, a two dimensional crystal with honeycomb lattice structure, became
recently available in research laboratories, a new variety of hybrid structures has been explored.
This lead to the development of promising technological devices and the understanding of the non-intuitive
physical mechanisms of relativistic-like massless fermions.
It is well-known that electrons propagating through single layer graphene (SLG)  exhibit many
peculiar properties, e.g. a gapless linear dispersion around the neutrality (or Dirac) point which
resemble massless Dirac fermions, a minimum conductivity at
zero carrier concentration, high mobility as well as the absence of backscattering.
When in contact to superconductors, graphene exhibits exotic superconducting properties.
Although graphene was not found to sustain intrinsic superconductivity itself, there is experimental
evidence
\cite{kanda_dependence_2010,tomori_fabrication_2010,heersche_bipolar_2007,du_josephson_2008,
ojeda-aristizabal_tuning_2009} that when
in proximity with a conventional superconductor it becomes superconducting.
Because of the conventional superconducting proximity effect, which describes how Cooper pairs
diffuse from the superconducting
material into metals, superconducting hybrid structures like superconducting-normal-superconducting
(SNS) Josephson junctions rise as an interesting systems in which one can study electronic
correlation of relativistic-like particles.
These experiments have attracted considerable theoretical attention insofar as it was predicted
that a finite supercurrent should exist even at zero doping where the density of states is vanishing
\cite{titov_josephson_2006} or that specular Andreev reflections should happen at the
superconductor/graphene interface in SLG \cite{beenakker_specular_2006}.
In the ballistic regime, which is realizable in SNS graphene Josephson
junctions according to recent experiments
\cite{du_josephson_2008,miao_phase-coherent_2007}, theoretical studies have shown
the existence of a finite bipolar superconducting current through the junction
\cite{titov_josephson_2006, hagymasi_josephson_2010, alidoust_tunable_2011,
halterman_characteristic_2011, black-schaffer_self-consistent_2008}.
%

%
With a gapless parabolic, instead of linear, band structure, bilayer graphene (BLG) appears at this
point as a suitable alternative for investigating electronic correlations in two dimensional
systems.
Also, as it is widely known theoretically and experimentally, a tunable gap can be  induced
in BLG by an out-of-plane applied electric field, which is very useful for transistor
applications \cite{castro_neto_electronic_2009,castro_biased_2007}.
We expect that the role of the gapless parabolic dispersion may be important since the
superconducting correlations depend strongly on the electronic properties of the material.
Moreover, the ability to open a gap in the spectrum by an external electric field could be of interest in  superconducting
devices.

Within a tight-binding Bogoliubov-de Gennes formalism we calculated self-consistently  the pair
correlation and the Josephson current
through a Superconducting-BLG-Superconducting Josephson junction. 
Our findings resemble previous numerical results shown for SLG-based Josephson junctions except
near the Dirac point in case of a long junction where a suppression of the current is found, mainly
due to the vanishing density of states present only in non-dimer sites.
We also show how the superconducting current can be switched off by applying a out-of-plane electric
field.

\section{Model}
We consider bilayer graphene in the common AB (Bernal) stacking with two inequivalent sublattices
\textit{A} and \textit{B} in the
top layer, and the corresponding sublattices \textit{\~{B}} and \textit{\~{A}} in the bottom layer. We model the interlayer
interaction by a hopping parameter which couples the nearest-neighbors in sub-lattices \textit{A} and \textit{\~{B}} from the top
and
bottom layer. Any additional interlayer coupling terms are ignored in this
study. It is well known that in this case the electron
dispersion displays two parabolic bands touching at the Dirac point and two
additional parabolic bands at higher energies due to interlayer induced
splitting \cite{guinea_electronic_2006,castro_neto_electronic_2009}.

\begin{figure}[ttt]
\includegraphics[width=\columnwidth]{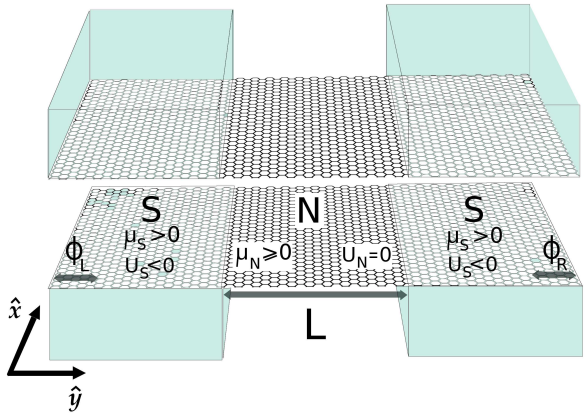}
\caption{\label{expsetup} Layout of the SNS-BLG Josephson junction where the superconducting leads are
modeled by assuming on-site attractive pairing potential $U_S$ and a heavy doping $\mu_S$ in the regions under the contacts
labeled
by \textit{S}. For the  non-superconducting link, with length $L$ and labeled by \textit{N}, we took the pairing potential
$U_N=0$ and a varying chemical potential $\mu_N$. Phases $\phi_L$ and $\phi_R$ are kept fixed during the self-consistently
calculation.}
\end{figure}

We study the Josephson effect in a BLG-based junction by considering the hybrid nanostructure illustrated in
Fig.~\ref{expsetup}.
The top and bottom layers are both in contact with superconducting leads while a junction of size \textit{L} is suspended.
The influence of the superconducting contacts is modeled  by assuming an on-site attractive pairing potential, $U_S<0$ and
heavy doping, $\mu_S>0$ in the contact regions, which are labeled by \textit{S}, such that a s-wave
superconducting state is induced in the outermost regions of both layers.
The normal region labeled by \textit{N}, which  has a tunable Fermi level $\mu_N$ and zero pairing potential, $U_N=0$, acts like a
non-superconducting channel through which Cooper pairs could tunnel.
Similar models for graphene based Josephson junctions were previously considered in the Dirac limit
\cite{titov_josephson_2006,alidoust_tunable_2011,hagymasi_josephson_2010,halterman_characteristic_2011} as well as in the
tight-binding formulation
\cite{black-schaffer_self-consistent_2008,black-schaffer_strongly_2010,covaci_superconducting_2011}.
We solve self-consistently for the order parameter along the junction following closely the self-consistent
calculation performed in Ref. \onlinecite{black-schaffer_self-consistent_2008}
for a ballistic single-layer graphene Josephson junction.
This is necessary in order to consider the possibility of Cooper pairs being
depleted close to the interface in the superconducting region due to the
existence of the normal region, i.e. the inverse proximity effect.
As is usually assumed for SLG-based junctions we consider a clean and smooth interface such that all
physical quantities
are homogeneous along the $\hat{x}$-direction parallel to the interface. In addition, we have
considered a wide junction, with width
$W>>L$,  as well as periodic boundary conditions imposed along the $\hat{y}$-direction.
The latter assumptions make it possible to reduce the self-consistent
calculation of the order parameter from a two-dimensional problem to a
one-dimensional
 one since we can restrict the calculation to only a unit cell along the
$\hat{y}$-direction perpendicular to the interface.
It is worth mentioning that the pairing potential does not distinguish between
sublattices defined on either layer. However, due to the interlayer coupling,
the self-consistent calculation should be performed
separately for only two inequivalent sites \textit{A} and \textit{B} (or \textit{\~A} and \textit{\~B}) within the unit cell.
The breaking of the inversion symmetry, which is easily achieved by considering a potential difference
applied between the layers, makes it necessary
to perform the self-consistent calculation for all four sub-lattice types within the unit cell.

In order to describe the DC Josephson effect we fix the difference in the phases
of the order parameter, $\Delta\phi^S=\phi_R-\phi_L$, in the outermost parts
of the \textit{S} regions (see Fig.
\ref{expsetup}). In a finite region near the interfaces, with size of the order of
the coherence length, the order parameter is allowed to relax
self-consistently. Therefore the phase gradient over the non-superconducting
region $\Delta\phi^N$ will be restricted according to
$\Delta\phi^N\leq\Delta \phi^S\leq\pi$, considering that the maximum value of  ${\Delta\phi^S}$ is
$\pi$.
This constraint for $\Delta\phi^N$ was recently pointed out by Black-Shaffer \textit{et al.}
\cite{black-schaffer_self-consistent_2008} in a SLG-based ballistic junction.

%
%

%
\section{Numerical Method: Chebyshev expansion of the Green's function}
Superconducting correlations in a BLG-based Josephson junction are described by
using the following tight-binding Hamiltonian with on-site
attractive Hubbard interactions:
\begin{eqnarray}
\nonumber
\ham&=&-\sum_{ \la i,j \ra \sigma} t\left( c_{i\sigma}^\dagger c_{j\sigma} + \tilde{c}_{i\sigma}^\dagger \tilde{c}_{j\sigma} \right) -
t_{\bot}  c_{i\sigma}^\dagger \tilde{c}_{j\sigma} \\
\label{ham1}
 &-& \sum_{i\sigma} \left( \mu_{i}+\epsilon_1 \right) c_{i\sigma}^\dagger c_{i\sigma} + \left( \mu_{i}+\epsilon_2 \right)\tilde{c}_{i\sigma}^\dagger \tilde{c}_{i\sigma} \\
\nonumber
&+&\sum_{i}U_i \left( c_{i\up}^{\dagger}c_{i\up}c_{i\down}^{\dagger} c_{i\down} + \tilde{c}_{i\up}^{\dagger}\tilde{c}_{i\up} \tilde{c}_{i\down}^{\dagger} \tilde{c}_{i\down} \right)
\end{eqnarray}
where $c_{i\up}^{\dagger}|$vac$\ra$ creates a spin-up electron  on the $i$-site in the top layer whereas
$\tilde{c}_{j\down}|$vac$\ra$ creates a spin-down hole on the $j$-site in the bottom layer.
The hopping parameter, $t$, describes the intralayer hopping integral between next-nearest neighbors in the same layers while
$t_{\bot}=0.143t$ correspond to the interlayer nearest-neighbors hopping which couples the dimer sites \textit{A} and
\textit{\~B}. Other hopping terms, like the interlayer coupling between the non-dimer \textit{B} and \textit{\~A} sites, are not
considered in the present work, since they influence only very low-energy excitations.
The Fermi level is shifted from the charge neutrality point or Dirac point by
the chemical potential $\mu_{i}$  and $U_{i}$ is the on-site
attractive pairing potential which is non vanishing only in the right and left
superconducting regions.
The on-site energies  $\epsilon_1$ and $\epsilon_2$  for atomic sites on the top and bottom layer, respectively,  have been
introduced in order to simulate a potential difference or gate voltage $V_g=\epsilon_1-\epsilon_2$ between the layers.

By using the Hartree-Fock decomposition and keeping only terms relevant to the
superconducting order, one can transform the many-body Hamiltonian
(\ref{ham1}) into a mean-field single-particle Hamiltonian, which within the
Nambu formalism can be written as follows:
\begin{eqnarray}
\label{ham2}
\ham=\sum_{\la i,j \ra} \vech{c_{i\up}^\dagger}{\tilde{c}_{i\up}^{\dagger}}{c_{i\down}}{\tilde{c}_{i\down}}
\mat{\hat{\ham_0}}{\hat{\Delta}}{\hat{\Delta}^{\dagger}}{-\hat{\ham}_0^{\dagger}} \vecv{c_{i\up}}{\tilde{c}_{i\up}}{c_{i\down}^\dagger}{\tilde{c}_{i\down}^\dagger}
\end{eqnarray}
where $\hat{\ham}_{0}$ and $\hat{\Delta}$ are the following  $2\times 2 $ matrices:
\begin{eqnarray}
\label{ham0}
\hat{\ham}_{0}=\mat{\epsilon_1+\mu_i}{0}{0}{\epsilon_2+\mu_i}(-\delta_{ij}) -
\mat{t}{t_{\bot}}{t_{\bot}^{\ast}}{t^{\ast}}(1-\delta_{ij}),
\end{eqnarray}
\begin{eqnarray}
\label{del}
\hat{\Delta}=\mat{\Delta_i}{0}{0}{\tilde{\Delta}_i}\delta_{ij}
\end{eqnarray}
where the diagonal elements of the matrix (\ref{del}) correspond to the on-site  mean-field
superconducting order parameter
$\Delta_i=U_i\la c_{i\up} c_{i\down} \ra$.
%

Following Refs. \onlinecite{covaci_efficient_2010} and \onlinecite{weise_kernel_2006} we have performed the self-consistent
mean-field calculation through a numerical approximation of the Gorkov Green's function by using the
Chebyshev-Bogoliubov-de-Gennes method. Both, the normal and anomalous Gorkov Green's function, can be approximated by a
superposition of a finite number of Chebyshev polynomials as follows:
\begin{eqnarray}
\label{greenseries}
\bar{G}_{ij}^{1\alpha}(\tilde{\omega})=\frac{-2i}{\sqrt{1-\tilde{\omega}^2}}\left[\sum_{n=0}^N
a_n^{1\alpha}(i,j)e^{-in\arccos(\tilde{\omega})}\right],
\end{eqnarray}
where the expansion coefficients for the diagonal, or normal ($\alpha=1$), and
the off-diagonal, or anomalous ($\alpha=2$), components of
the 2$\times$2 Green function are defined respectively as
\cite{covaci_efficient_2010}:
\begin{eqnarray}
\label{normal}
a^{11}_{ij}(\omega) = \la c_{i\up}\left|T_n(\ham)\right|c_{j\up}^{\dagger}\ra \\
\label{anomalous}
a^{12}_{ij}(\omega) = \la c_{i\down}^{\dagger}\left|T_n(\ham)\right|c_{j\up}^{\dagger}\ra^{\ast}
\end{eqnarray}
where $T_n(x)=acos(n \cos(x))$ is the Chebyshev polynomial of order $n$, which is defined according
to the recurrence relation: $T_{n+1}(x)=2xT(x)-T_{n-1}(x)$.

In order to perform the sum  (\ref{greenseries}) one needs first to rescale the Hamiltonian (\ref{ham2}) such that
the  eigenvalues lie in the [-1,1] interval.
To this end,  matrices (\ref{ham2}) and (\ref{del}) as well as the energies have being
normalized according to:
$\ham\rightarrow \tilde{\ham}=(\ham-\eins b)/a$
and $\omega\rightarrow \tilde{\omega}=(\omega-\eins b)/a$, where the rescaling
factors are $a=(E_{max}-E_{min})/(2-\eta)$ and $b=(E_{max}+E_{min})/2$,
with $\eta>0$ being a small number.

Once the Hamiltonian is normalized, the expansion coefficients can be obtained through a recursive procedure.
Starting with an initial vector $|j_0\ra=|c_{j\up}^{\dagger}\ra$ and  the first
term in the iteration  $|j_1\ra=\ham |j_0\ra$ one can obtain the $n$-th
term by using the Chebyshev recurrence relation:
$|j_n\ra=2\ham|j_{n-1}\ra-|j_{n-2}\ra$.
Chebyshev moments are finally obtained from the scalar product $\la \alpha|j_n\ra$, where $\la \alpha|$ are the
vectors $\la 1|=\la c_{i\up}|$ and $\la 2|=\la c_{i\down}^{\dagger}|$ for
moments expanding the (\ref{normal}) and (\ref{anomalous}) component of
the Green's function respectively.

Since most of the computational effort corresponds to  sparse matrix-vector and vector-vector
multiplication,  high speed-up  can be achieved by implementing parallel
computation on graphical processing units (GPUs). We are therefore able
to solve efficiently systems described by matrices of sizes  between
88K$\times$88K and  320K$\times$320K according to the junction size considered
in this study.
Additional parallel computations can be implement by considering that all physically quantities calculated here, such as
density of states, pair correlation or the Josephson current, can be obtained from the Green's
function for each lattice point separately.

\section{Results}
It is well-known that the energy dispersion in SLG and BLG differ  around
the neutrality point. Therefore, qualitative   differences are expected in the
proximity effect as well as in the Josephson current in the two systems.
For a quantitative comparison between SLG and BLG Josephson junction we have set up the following values for the physical
input parameters: $U_S=-1.36t=-3.4$eV and $\mu_S=0.6t=1.5$eV, similar to
the values used in a previous self-consistent study for ballistic SLG Josephson  junctions
\cite{black-schaffer_self-consistent_2008}. These parameters lead to a finite s-wave bulk superconducting
order parameter,  $\Delta_0=0.041t$, which corresponds to a superconducting coherence length $\xi=\hbar v_F/\Delta_0 \approx 23$ 
unit cells similar to the one considered in the SLG junction case. Due to
the difference in the local density of states between dimer and non-dimer locations, the order parameter is slightly  different
for these two types of atoms. Both junction length regimes are solved
with the proposed self-consistent numerical method, e.g.  short junction for $L<\xi$ and long junction for  $L>\xi$.

Previous analytical descriptions of BLG Josephson junctions based on the Dirac
equation \cite{alidoust_tunable_2011,snyman_ballistic_2007} requires
smooth interfaces and a low energy regime, for which
$\Delta_0<<t_{\bot}<<\mu_S$.  Here these restrictions are lifted but in order to
compare with relevant experimental scenarios we performed calculations only for situations
corresponding to $\Delta_0<t_{\bot}<\mu_S$.

\subsection{Proximity effect}
\begin{figure}[ttt]
\includegraphics[width=\columnwidth]{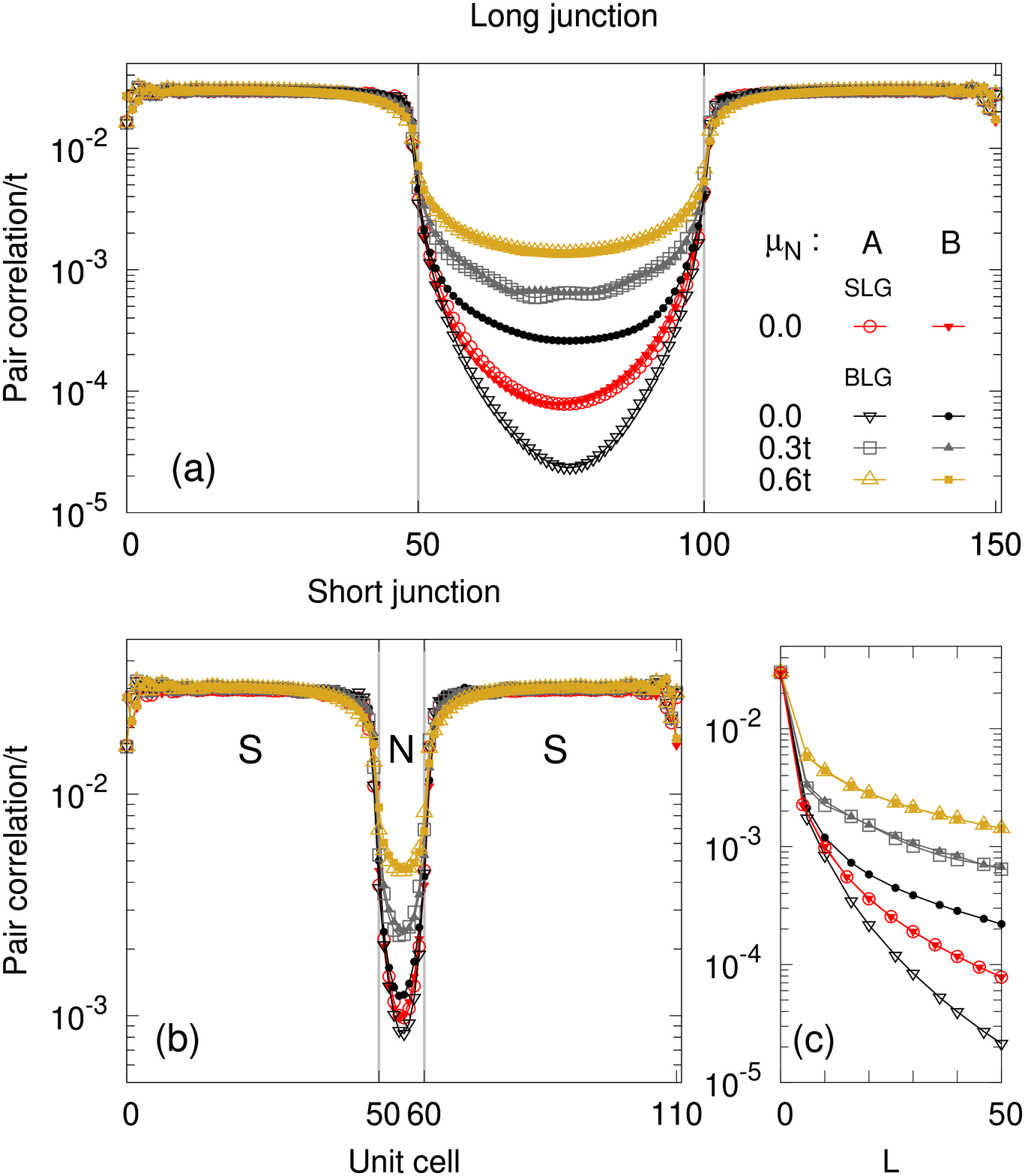}
\caption{\label{pairamp} Absolute value of the pair correlation in a long (a) and short (b) BLG Josephson junction  as a
function of the position along the $\hat{y}$-direction perpendicular to the \textit{SN} interface.
Both inequivalent sites, dimer at \textit{A} and non-dimer at \textit{B}, are plotted separately for different doping levels
considered for the non-superconducting region: $\mu_N=0$, $0.3t$ and $0.6t$. The
first one correspond to the case when the Fermi level is pinned at the Dirac
point while the last  one corresponds to no-FLM at the interface.
(c) Pair correlation at $L/2$ in the \textit{N} region as a function of the junction length $L$ for
different values of $\mu_N$. SLG self-consistent results are shown for comparison with the BLG undoped case.
 }
\end{figure}

We show in Fig. \ref{pairamp} the self-consistently calculated
pair correlation $\la c_{i\up} c_{i\down} \ra$ for both inequivalent sites, \textit{A} (dimer) and \textit{B} (non-dimer), in a
unit cell defined along the $\hat{y}$-direction perpendicular to the interface.
As was previously mentioned, both long and short junctions are considered here and plotted in
Figs. \ref{pairamp}(a) and \ref{pairamp}(b), respectively.
We present the pair correlation function for several doping levels in the
non-superconducting region showing that the proximity effect is strongly
dependent on the relative Fermi level mismatch between the \textit{S} and
\textit{N} regions.
A significant difference between BLG and SLG can be seen for the undoped case
($\mu_N=0$) while for other dopings the pair correlation is found to exhibit
similar behavior in BLG and SLG.
%
%
In particular, we can see that in the undoped case dimer sites in BLG show a suppression of the pair correlation over the
\textit{N} region compared to the SLG case. Opposite behavior is seen for non-dimer sites where a larger pair leakage into
\textit{N}
is found. This behavior is similar to the proximity effect in strained graphene where a sublattice polarization of the local
density of states in the zero-th pseudo-Landau level induces sub-lattice dependent leaking distances
\cite{covaci_superconducting_2011}.
No relevant differences in the pair correlation profile are found  between
dimer and non-dimer sites in BLG for higher doping levels considered here:
moderately doped $\mu_N=0.3t$ and highly doped $\mu_N=0.6t$.
Note also in Fig. \ref{pairamp}(b) that the reported interlayer asymmetry in the pair correlation in BLG is found 
to be not important for the short junction regime.
In fact, we can clearly see in Fig. \ref{pairamp}(c) that the difference between
the pair correlation at sites \textit{A} and \textit{B} becomes larger
as the junction length is increased, as large as a few orders of magnitude.
\subsection{LDOS}
\begin{figure}[ttt]
\includegraphics[width=\columnwidth]{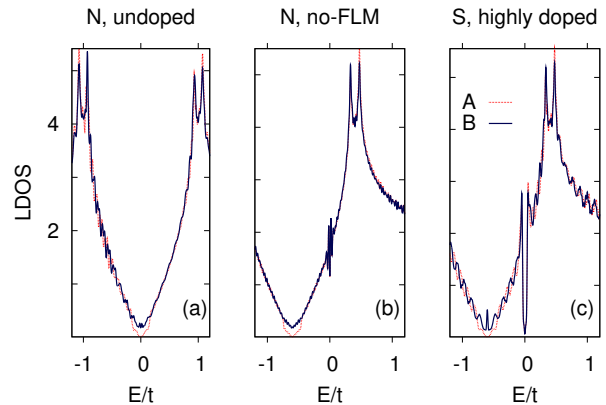}
\caption{\label{ldos} LDOS for A (red line) and B (blue dashed line) sub-lattice
sites in the \textit{N} region with  (a) zero doping (at the Dirac point)
and (b) heavy doping (or with no-FLM between the superconducting and normal
regions). (c) LDOS in the highly doped \textit{S} region showing the
superconducting  gap.
}
\end{figure}

To further understand this peculiar behavior observed only in the undoped case
we have plotted the local density of states (LDOS) in the superconducting
and non-superconducting regions. The LDOS is plotted for dimer (\textit{A}) and
non-dimer (\textit{B}) lattice sites along the inhomogeneous direction in the BLG
Josephson junction away from the \textit{SN} interface.
Three particular cases have been chosen and shown in Fig. \ref{ldos}. Panel (a) shows the LDOS for two sites in the
\textit{N} region for the undoped case, whereas, in panel (b) we have depicted LDOS for the same location but when there is
no-FLM.
Results shown in Fig. \ref{ldos}(a)-(b) are  consistent with the fact that the density of states at dimer sites
vanishes linearly around the Dirac point, while being finite at the non-dimer
sites\cite{guinea_electronic_2006,castro_neto_electronic_2009}.

Due to the differences in the LDOS for inequivalent sites in BLG the proximity
effect will give different leaking distances in different sub-lattices,
as we show in Fig. \ref{pairamp}(c) for the undoped case.
Increasing the doping level in the normal region leads to the LDOS becoming homogeneous. This result is consistent with
the fact that no difference between both kind of sites is observed for the other doped cases in the pair correlation function
shown in the last section.
Finally, panel (c) in Fig. \ref{ldos} shows the LDOS deep inside the
superconducting region where we can see clearly the coherence peaks on each side
of
the superconducting gap. Note that in the normal region, the LDOS is modified
near the Fermi level, corresponding to the formation of Andreev levels.
\subsection{Josephson current}
\subsubsection{Unbiased case}
\begin{figure}[ttt]
\includegraphics[width=\columnwidth]{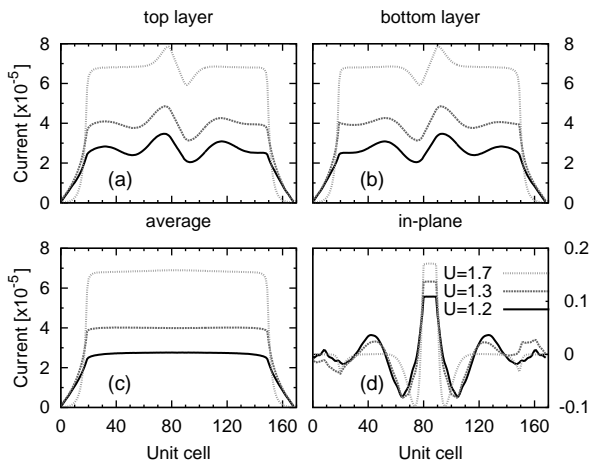}
\caption{\label{currpro} Self-consistent Josephson current as a function of position along the
$\hat{y}$-direction for top (a) and bottom (b) layer as well as the average
current (c) in the system for different values of the pairing potential,
i.e. different coherence lengths. The interlayer current between the
\textit{A}-\textit{\~B} dimer sites is plotted in (d).}
\end{figure}

\begin{figure}[ttt]
\includegraphics[width=\columnwidth]{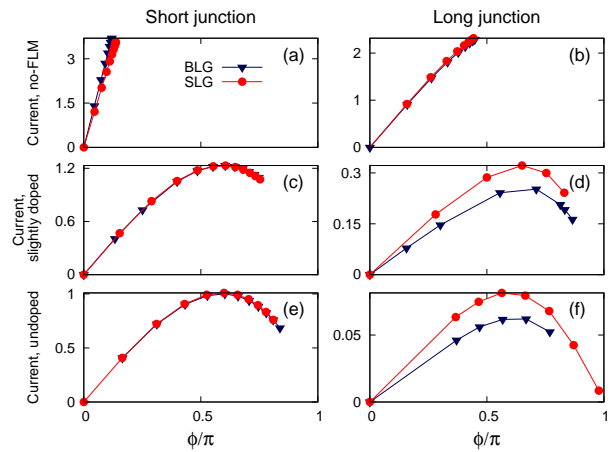}
\caption{\label{cpr} Current-phase relation normalized to the critical current found in (e) for SLG and BLG-based junctions
considering  different doping levels in the \textit{N} region: (e),(f) undoped $\mu_N$=0, (c),(d) slightly doped $\mu_N$=0.1$t$
and (a),(b) no-FLM $\mu_N$=$\mu_S$=0.6$t$ for  short (a),(c),(e) and long (b),(d),(f) junctions. The short and long junction lengths are $L=10$ and $L=50$, respectively.
    }
\end{figure}

In the absence of applied bias voltage but in the presence of a finite phase
difference between the two superconducting sides,  a DC  supercurrent will
flow across the junction
\cite{black-schaffer_self-consistent_2008,covaci_proximity_2006}. This is the
usual DC Josephson effect.
For this purpose a phase bias is achieved by fixing a desired phase difference
between the outermost parts of the superconducting regions,
 $\phi_L$ and $\phi_R$ for the left and right sides of the junction respectively
(see Fig.~\ref{expsetup}).
In order to numerically calculate the Josephson current we solve self-consistently for both phase and amplitude of the order
parameter along the junction except in the extreme regions where we keep the phases fixed.
The current profile along the junctions in both layers as well as the interlayer
current are shown in Figs.~\ref{currpro}(a)-(d) for an undoped
non-superconducting link for different values of the pairing potential.
As we can see, the supercurrent is found not to be constant within each layer separately, contrary
to what is usually expected to happen for self-consistent
current calculation in 2-dimensional systems \cite{covaci_proximity_2006}.
Instead, one interesting feature in BLG is the appearance of a weak interlayer current
between \textit{A}-\textit{\~B} dimer
sites as a consequence of the current conservation law.
We observe that the LDOS at the left interface is asymmetric in top-bottom layers while for the right interface  the asymmetry is
reversed.
%
%
Because of this, the current is enhanced at the left interface in the top layer
while being suppressed in the bottom layer, therefore a weak
interlayer current appears. The reverse happens at the right interface, but
the average current remains flat across the whole junction, as expected.
%
%
Next we construct the current-phase relation (CPR) by performing self-consistent current
calculations for different phase differences between the superconducting contacts and doping levels
in the non-superconducting region.
We show in Fig.~\ref{cpr} the phase-dependence of the current for both SLG and
BLG for no-FLM cases with $\mu_N=0.6t$, panels (a) and (b), slightly doped with $\mu_N=0.1t < t_{\bot}$	, panels (b) and (c)
and undoped, panels (e) and (f) considering short, panels (a)-(c)-(e), and long junction, panels (b)-(d)-(f), regimes.
Note that, in most of the cases shown in Fig.~\ref{cpr} a complete description of the current over
the full [0,$\pi$] phase-range is not possible.
This restriction appears as a consequence of the relaxation of the phase over the \textit{S} region which becomes significant as
the FLM is reduced in the junction until \textit{S} and \textit{N} regions are equally doped and the phase drop goes linearly
through
the self-consistent region. Therefore the phase difference over the normal
region will always be smaller than the applied phase difference in the
 superconducting regions.
A similar constraint is found for SLG-based junctions, as is shown in
Fig.~\ref{cpr} and which was previously pointed out by Black-Schaffer \textit{et al.}
\cite{black-schaffer_self-consistent_2008}.
We find that for no-FLM situations the Josephson
current density and the CPR of the BLG is almost identical to the one in SLG for both short and long junction.
In contrast, for the undoped and slightly doped cases the short and long junctions have different
behavior. Short BLG junctions are similar to short SLG ones, while for long BLG
junctions the Josephson current is suppressed. The origins of this suppression
could be traced back to the sublattice polarization of the leaking distance  which remains even for doping levels
lower than the interlayer hopping energy, $t_{\bot}$.
While one sublattice (non-dimer sites) has an enhanced leaking distance, the
other sublattice (dimer sites) behaves like an insulator with a short leaking
distance. The resulting combination corresponds to a slightly suppressed Josephson current when compared to SLG.

\subsubsection{Biased case}
Inversion symmetry can be broken in the BLG nanostructure by considering a potential difference, $V_g=\epsilon_1-\epsilon_2$
with $\epsilon_1=V_g/2$ and $\epsilon_2=-V_g/2$,  applied  between the layers.
As a consequence, a tunable gap $\Delta_0$ is induced at the Dirac point for the undoped case and therefore inversion symmetry
breaking appears as a good possibility to switch off the superconducting current when the voltage
induced gap overtakes the superconducting gap, $\Delta_g > \Delta_0$.
At this point, we only  consider short junctions  where such proposal might have potential
technological applications.
\begin{figure}[ttt]
\includegraphics[width=\columnwidth]{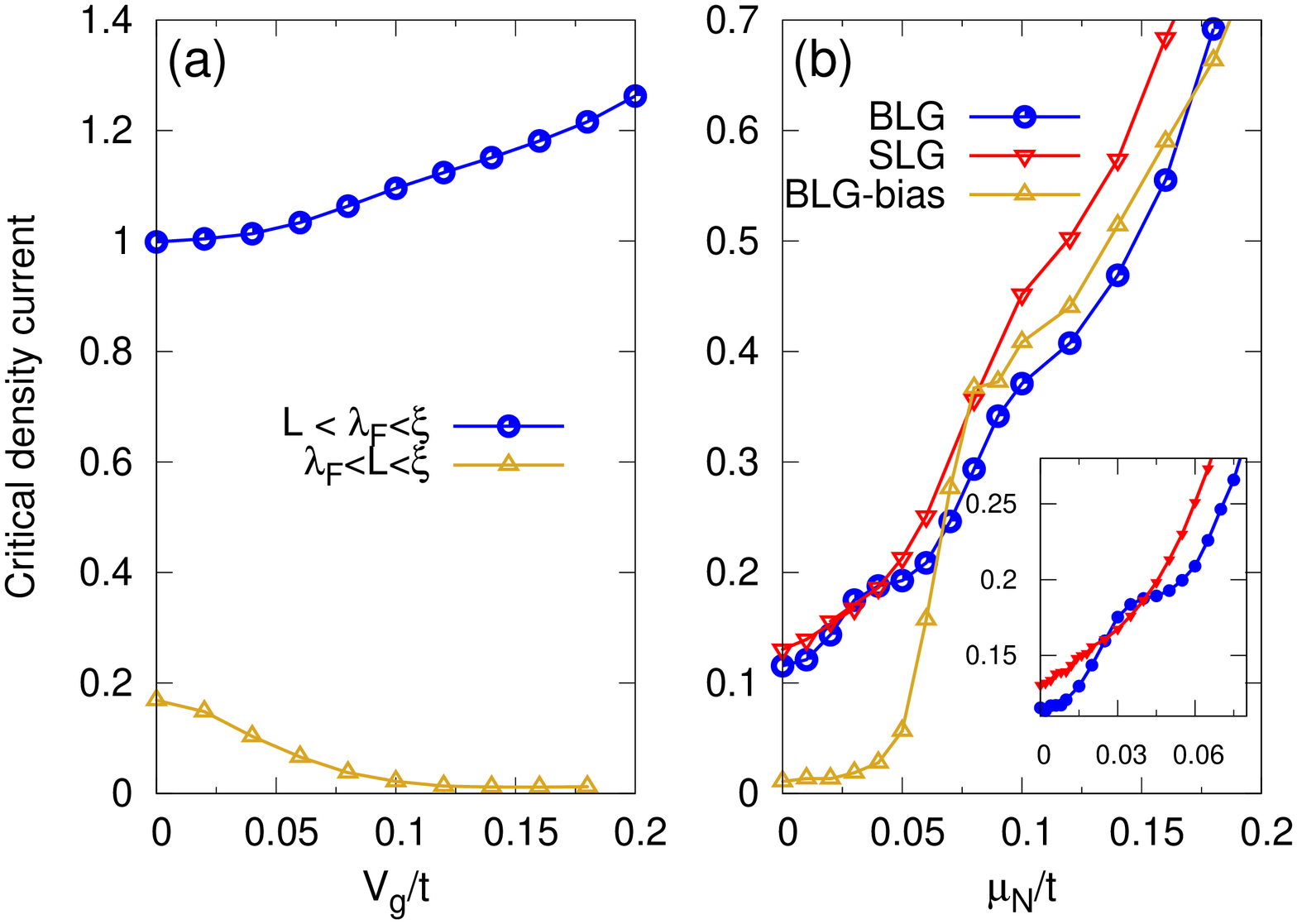}
\caption{\label{curr} Critical density current in a short BLG junction, where $L=40$ Unit cells considering a lower pairing
potential $U_S=-1.2t$, as a function of the gate voltage $V_g$ at the Dirac point (a) and the chemical potential $\mu_N$ (b). In
the left figure two different regimes are considered: $L<\lambda_F$ and $L>\lambda_F$, showing an enhancement and
suppression of the superconducting current, respectively. On right side $\mu$-dependence is plotted for bias and unbias cases. SLG
is included for comparison.
    }
\end{figure}
In Fig.~\ref{curr}(a) we show the dependence of the supercurrent on the applied bias voltage,
$V_g$, perpendicular to the BLG layer. We observe two distinct
regimes, according to the relation between the Fermi wave-length and the junction length. 
In the case  when $\xi>\lambda_F=\hbar v_F/(V_g/2)>L$, we observe an enhancement of the Josephson current. This
is because the charge density in the junction does not recover its bulk expected value when the BLG is  under bias ($+n$ for one
layer and $-n$ in the other layer) but has a finite positive value. 
In this regime the junction is effectively doped, thus showing an enhancement of the current. A different  dependence of the
current is achieved when $\xi>L>\lambda_F$. In this case for the center of the
junction the charge density has the opposite polarity, and the expected suppression of the current with bias  voltage is obtained.

In Fig.~\ref{curr}(b) we show the $\mu$-dependence of the current for the biased and unbiased cases.  A doping activation is found
in the bias case for an energy value around $\Delta_g$,
in addition a slight increase in the current is observed at $\mu_N \approx \Delta_g$ due to the  enhancement of the LDOS above the
gap edge.
SLG results are included in order to compare directly with BLG and show that for the chosen length the current density is slightly larger in SLG. In the inset of Fig.~\ref{curr}(b) we focus on the low doping regime and observe that in BLG additional oscillations appear, reminiscent of what was previously found analytically in the Dirac approximation \cite{alidoust_tunable_2011}.

\section{Conclusions}
In conclusion by using an efficient numerical procedure we solve self-consistently the Bogoliubov de Gennes equations for a tight-binding model of the AB-stacked bilayer graphene Josephson junction.
When compared to single layer graphene Josephson junctions we uncover several regimes. First, in the short junction regime, the
current density is similar for SLG and BLG for any doping
of the normal junction region. Second, in the long junction regime, for undoped junctions, the BLG current density is suppressed while for doped junctions (with doping larger than $t_\perp$) the BLG
and SLG junction behave in a similar way. We attribute the peculiar behavior of the undoped BLG junctions to the difference of the
LDOS between the dimer and non-dimer sites, which give suppressed or
enhanced Cooper pair leaking distances depending on the sublattice. We have calculated the current-phase relation and showed that
similar to SLG, even for short junctions there is a departure from
conventional symmetric CPR. Finally, we have shown that by applying a gate voltage perpendicular to the BLG a gap in the spectrum
can be induced and a supercurrent switch can be achieved given that the
junction length is larger than the Fermi wave-length.
\section*{Acknowledgements}
This work was supported by the Flemish Science Foundation (FWO-Vl).


%

\end{document}